# THE DYNAMIC RELATIONSHIP OF CRUDE OIL PRICES ON MACROECONOMIC VARIABLES IN GHANA: A TIME SERIES ANALYSIS APPROACH

**Dennis Arku**

Email: darku@ug.edu.gh

Department of Statistics and Actuarial Science,

College of Basic and Applied Sciences,

University of Ghana, Legon**.**

**Gabriel Kallah-Dagadu**

Email: gkallah-dagadu@ug.edu.gh

Department of Statistics and Actuarial Science,

College of Basic and Applied Sciences,

University of Ghana, Legon.

**Dzidzor Kwabla Klogo**

Email: dkklogo@gmail.com

Department of Statistics and Actuarial Science,

College of Basic and Applied Sciences,

University of Ghana, Legon.

**Corresponding Author: Gabriel Kallah-Dagadu** (gkallah-dagadu@ug.edu.gh)

**Abstract**

The study investigates the effects of crude oil prices on inflation and interest rate in Ghana using data obtained from Bank of Ghana data repository. The Augmented Dickey-Fuller and the Phillips-Perron tests were used to test the presence or otherwise of unit root relationship among the variables. The stationarity test showed that the variables are either integrated of order one or integrated of order zero. Autoregressive distributed lag bounds test approach was adopted to examine cointegration among the variables. The results showed a positive relationship between crude oil prices and inflation in the long-run. The short-run show the coefficient of the first period lag value of inflation is negative, but statistically insignificant in the short-run. However, the second period lag value of inflation is positive and significant. The result also shows negative relationship between crude oil price and interest rate. Based on the findings, it is recommended that the government of Ghana should provide and strengthen the efficiency of the public transport system to help reduce transport fares in order to shield the poor from the implications of oil price increases in Ghana.

## 1.  INTRODUCTION

Foreign currencies are the monetary value that are normally used in quoting oil prices. This is because hydrocarbons are deemed international commodities with worldwide demand and supply forces.  This makes it easier for the management of disparities among domestic currencies and external currencies. (Abouali, Nikbakht, Ahmadi, & Saadabadi, 2009). Crude oil prices have been recognized as contributing to exchange rate depreciation, particularly in net crude oil importing nations (Gatsi, Gadzo, Anipa, & Kosipa, 2015).

Ghana, a producer of petroleum is heavily engaged in petroleum exports and imports and it has affected the economy due to the fluctuation in crude oil prices with the controlled float scheme. Moreover, the slowdown in the cost of crude oil since 2014 has decreased its contribution to petroleum-related revenue in Ghana, formerly representing 21.1 percent of the government budget (Bawumia & Halland, 2017).

Macroeconomic volatility implies that shocks are susceptible to macroeconomic variables. In terms of shock resilience, macroeconomic variables such as GDP, inflation, currency, interest rates and others tend to be volatile and low in restraining shock. It is a situation in which little economic shock is subject to macro-economic changes and uncertainty. In view of this, many researchers have investigated the impact of oil price modifications on macroeconomic variables. The agreement they have reached is that, while changes in the petroleum prices are directly associated with many macroeconomic factors, they do not significantly affect production development (Ewing & Malik, 2017; Taiwo, Abayomi & Damilare, 2012).

A change in the parameters of the price of petroleum may or may not involve a shift to the fundamental principles of any economy. Rising petroleum prices have a beneficial impact on petroleum exporters' macroeconomic development, while lower prices have the opposite impact (Saint-Georges, 2002). At the macro level, financial activities have shown to be more influenced by petroleum price rises than declines (Jacobs, Kuper, & Van Soest, 2009).

The absence of complete empirical research on the effect of oil price on Macroeconomic factors in developing nations, such as Ghana, both exporting and importing oil is a deficiency of previous research or reviewed literature. Furthermore, the models that have been used do not capture the economic and financial time series vibrant behavior.

The research thus seeks to address these shortcomings through the use of Auto Regressive Distributed Lag model (ARDL). This is a model class that captures a number of time series data on a range of conventional temporal structures to examine the connection of oil price to macroeconomic factors in Ghana.

## 2. LITERATURE REVIEW

Bala & Chin (2018) used OLS MA technique to identify the relation between inventory prices and the impact on four developing markets; India, Russia, Brazil and China. The investigator used petroleum prices, exchange rates and shift values as explicative factors, but the findings were small and this demonstrated market inefficiency. He found that such economies are emerging in order to influence national variables more externally as regards petroleum prices and exchange rates. According to Kilian and Hicks (2013), strong growth in global petroleum prices and violent currency fluctuation are usually seen as variables that discourage economic growth. In the case of oil exporters and negative of oil importer nations it was presented that an increase in oil prices, all other equivalent things, should be deemed positive and the contrary.

In his study Foresti (2006) discovered that the causal connection between the prices of stocks and capital expenditure for India and Bangladesh is unidirectional, while the Wang (2010) investigates the relation between financial operations, inventory price and petroleum price in three economies: Russia, China and Japan; Unlike Russia, however, China and Japan have had no co-integrative connection between the factors.

The relationship between crude oil and inflation during 2002-2018 was positive in another research of Ahmed, Bhutto and Kalhoro (2019) on the component cost conduct of crude oil and its effect on Indian inflation. They claim that humanity uses the world's power resources in a manner that no other animal has ever done; now the world is ruled by the oil fuel economy. Although there are so many alternative fuel sources, they still have a tiny share of the current demand worldwide. Oil prices and inflation are often seen as linked within a structure of causes and effects. Inflation follows in the same direction as oil prices rise or decline. Their research disclosed a correlation between crude oil prices and inflation of 0.726 for Pearson and a favorable connection. They say that Indian petroleum prices have risen simultaneously with global petroleum prices steadily. Since the start of the 20th century, oil prices have begun to rise substantially. The impact of the petroleum price shock is the higher manufacturing costs owing to the higher cost of fuel. The price of manufacturing would also increase and in turn lead to a reduction of supply due to economic inflation. They found that a substantial proportion of the decrease in the petroleum transmission worldwide is due to the decrease in exchange-rate inflation impact and the decrease in oil intensity. Their estimates however indicate that part of the decrease in oil passage is unexplained even when these factors have been checked.

In Zhihua, Meihua and Bo (2011), they analyzed the impact and delays of coal price fluctuation and the effect of the CPI on its effectiveness, taking into account the characteristics of coal energy production and use in the field of cointrusion, the ECM model, the pulse response function and other economic methods. From January 2002 to October 2010 China chosen information on the CPI and COP index to show the positive correlation between the fluctuation of coal prices and the CPI. They concluded that as the prices of coal fluctuate by 1%, the CPI will vary by 0,157 percent in other conditions. They claim that the model built by coal price and CPI will adapt to non-equivalence balancing states with the adjustment magnitude -0.03 in case of short-term fluctuation deviates from the long-term equilibrium.

The relations between petroleum prices and exchange rates are examined by Hussain, Zebende, Bashir and Donghong (2017), using different models, which include the two shorter- and long-term scales of 12 Asian nations. The adverse cross-relation coefficient is presented by India and China. The sample shows comparable outcomes in the majority of the nations with the exception of Honduran and Japan, which demonstrate a powerful beneficial cross-relation between the cost of oil and the exchange rate. The study's writers fail to understand the significance of national oil demand, which has serious price and inflation implications.

The import bills for oil are another key factor not provided sufficient attention. The effects of structural petroleum prices shock on inventory market return and volatility in the United States are investigated by Kang, Ratti and Yoon (2015). The study finds a strong significant covariance of structural oil shocks and stock return. The article refers to the research of the 2008 worldwide financial crisis and concludes that shocks related to the price of petroleum have a more significant effect on inventory returns than in previous years. But the writers did not take account of the other macroeconomic factors that could affect stock market yields.

Basnet and Upadhyaya (2015) analyzes in five ASEAN countries using the Structural Vector Autoregression (VAR) model the effect of the oil price shocks on real output, inflation, and real exchange rate. It was found that fluctuations in oil prices do not effect on the ASEAN markets in the long term using the cointegration method. The article notes that fluctuations in petroleum prices are not significantly influenced by big FDI influxes and big internal industry performance in macroeconomic performance. But, from the perspective of FDI and industry performance evaluation, this research ignores the effect of the oil prices on stock market yields.

In addition, Zhu, Li and Li (2014) are investigating the reliance of oil and inventory prices in ten Asia-Pacific nations. The findings show that the dependence between crude petroleum prices and the inventory market for Asia Pacific is very low with the use of AR (p) — GARCH (1,1) to model. Another key finding of this research is that, before the Hong Kong global financial crisis, there had been a favorable connection between oil and inventory prices. A key deduction from the research is that when the price of petroleum and inventory prices are highly dependent, a bigger amount of assets must go on inventory when the price of oil is increasing.

The Vector Auto Regression Approach (VARA) research was carried out in relation to the effect on the macro-economic performance of Nigeria by Omisakin and Olusegun (2008). The research found that oil price fluctuations had a significant impact on oil revenue and volatility in production. Therefore, the petroleum price shock has no significant impact on cash supply, price level or public spending in Nigeria during the period covered by the study. Maxwell (2013) have studied the effect of oil volatility on Nigeria's macroeconomic activity and found that the relationship between interest rates, exchange rates and oil prices is one-way. There was nevertheless no important connection between the volatility of petroleum prices and actual GDP. The document concludes that the volatility of petroleum prices is a significant factor in actual exchange prices, whilst the price volatility, rather than the volatility of petroleum prices, impacts production development in Nigeria.

Gay (2016) confirms the beneficial connection between rises in oil prices and the financial condition; it demonstrates the disruptive impacts in equilibrium of payments and public finances during periods of oil price declines. In addition, it was also said that even a tiny shock at world oil prices will have a long-term effect on economic growth in the country. In research on volatility of petroleum prices and Nigerian financial growth, Wilson, David, Inyiama and Beatrice (2014) have also investigated the causal link between petroleum prices and significant macroeconomic factors from 1980 to 2010. The results show that there is a favorable, but small link between oil prices and Nigerian GDP. In particular, oil prices have no significant impact on Nigeria's real GDP and exchange rate.

Kibunyi, Nzai and Wanjala (2018), conducting a survey in Kenya, have shown that the volatility of petroleum imports in Kenya has a major negative effect on GDP development and has had an impact on financial development The new runs as well as the long runs. Most study indicates, however, that a rise in petroleum imports leads to enhanced financial development and vice versa. The author could create only one co-integrating connection between the four studied factors (actual exchange rate, traffic volume, the overall production index and GDP growth rate) by using Joseph Johansen-Juselius's co-integration test strategy and vector error correcting method.

The empirical assessment of macroeconomic variables and crude oil prices in Ghana, undertaken by Gatsi, Gadzo, Anipa, and Kosipa (2015), was used to perform a time sequence and country-specific changes between 1980 and 2011 to analyze macroeconomic variables with regard to crude oil prices for Ghana. The Granger cause-effectiveness test performed a solid methodology. In general, the findings of the research revealed that GDPGR and the other macro-economic factors had important long-term relationships, such that LOP and REER long-term motions substantially explain GDPGR movement. The GDPGR and other macroeconomic factors were also considerably short-term in that shorter-term moves of world crude oil prices explained important motions in the GDPGR with shorter-term motions in the REER, explaining moves in the GDPGR. Acheampong (2007) has tried to predict the short-run and long-run impact of crude oil prices on a chosen macroeconomic variable: Ghana exchange rate, GDP, interest rate and inflation in an early survey in Ghana using Vector Error correction models. The research used information from 1970:1 to 2006:4 and found that the impact of crude oil was positive on inflation, while negative on production.

In Ghana also Cantah (2014) conducted a survey on the cost and economic development of crude crude oil. The ARDL method was used to examine, using annual information from 1967 to 2011, the connection between the cost of crude oil and the financial development of Ghana. This research has regulated the impact of fiscal policies on relations, contrary to prior research on Ghana's raw oil price financial growth relations. The findings showed that the price of crude oil and economic growth in Ghana are long-running. Moreover, the research shows that the increase in oil prices has an adverse effect on short- and long-term financial development, which has also been strengthened by rises in government expenditure in the form of fuel subsidies as a reaction to the cost of oil.

## 3. Materials and Methods

### 3.1 Materials

The study used secondary time series data sets. The data sets were monthly data and covered the period of 1/1/2000 to 31/12/ 2018. The data comprises of crude oil price (BCROIL), inflation and interest rate. Data on all variables were obtained from the Bank of Ghana (2018) and World Bank, World Development Indicators (WDI, 2018).

### 3.2 Model Specification

The research therefore examines the impact of the cost of crude oil on inflation in model 1 while considering the impact of crude oil prices on model 2 interest rates. The models are as follows:

$$INFL_t = f(OP_t) \qquad\qquad\qquad (3.1)$$

$$INT_t = f(OP_t) \qquad\qquad\qquad (3.2)$$

Where equation (3.1) is Model 1 and equation (3.2) is Model 2. This demonstrates the time variation nature of the factors and reflects inflation, interest and crude oil price respectives. Therefore, the estimated equation shape (3.1) and (3.2) is as follows:

$$\ln OP_t = \alpha_0 + \alpha_1 \ln INFL_t + \mu_t \qquad (3.3)$$

$$\ln OP_t = \delta_0 + \delta_1 \ln INT_t + \zeta_t \qquad (3.4)$$

## 3.3  Data Description

The description and measurement of the variables under consideration are presented in Table 1.0

**Table 1.0 Variable Definitions and Measurement**

| Variable | Notation | Definition | Measurement |
|---|---|---|---|
| Crude Oil Price | OP | The price of crude oil traded in the international market by oil producing countries. | The international crude oil prices (monthly) |
| Inflation Rate | INFL | Consumer price index - all items. Percentage change from previous period monthly | Monthly inflation rates |
| Interest Rate | INT | The rate at which loans are given to customers | Weighted average yearly lending rates |

## 3.4    Estimation Strategy

### 3.4.1 Unit Root

In order to avoid spurious outcomes, it is essential to examine the root characteristics of the unit of the variables used in the equation (3.2). For factors in time series that most probably contain root units, that is particularly crucial. The experiment determines the potential for any degree of permanence in a divergence if the variables are shocked.

### 3.4.1.1 Augmented Dickey-Fuller (ADF) Test

Compared with Dickey-Fuller (DF), this ADF is appropriate for complex and relatively tigger time-series data. The test increases the DF technique to guarantee white noise is produced without changing the distribution of test data under the null root hypothesis. The test is carried out in a steady way. The choice of delays is made by means of the Akaike Information Criterion (AIC). The experiment examines the null assumption that the root unit is not stationary, and therefore that no root unit is stable in the alternative hypothesis. The null hypothesis is dismissed if the computed statistical test is more than the ADF absolute critical value. The equation of ADF is:

$$\Delta y_t = \sigma + \gamma_t + \vartheta y_{t-1} + \sum_{t=1}^{k} \alpha \Delta y_{t-1} + \varepsilon_t \qquad (3.3)$$

The steady term and time trend coefficient are specified accordingly. The optimum lag duration is respectively the differential operator and equally and respectively are the error term and time variable.

### 3.4.1.2 Phillips-Perron (PP) Stationarity Test

The ADF technique is modified by a nonparametric technique. It can correct self-correlationship and error heteroscedasticity. This outcome is more robust than other assessment methods. The null and alternative hypotheses are similar to the ADF technique and are conducted in a steady and continuous way. The test is carried out both at and after differentiation. At rates, the stationary series is I(0) (i.e. zero[0]) while after first differentiation, I(1) (i.e. series incorporated in order one[1]) is stationary. The PP equation is: PP

$$\Delta y_{t-1} = \delta_o + \gamma y_{t-1} + \nu_t$$

(3.4)

### 3.4.2 Autoregressive Distributed Lag (ARDL) Model

In order to examine cointelation among the factors in the model, the research adopts the Autoregressive Distributed Lag (ARDL) test method. The research examines in particular the effect level of actual exchange rate, national gross product development, inflation, currency policy rate, and crude oil price interest rates. Under the Error Correction Model (ECM), Cointegration will be examined. For the period 2000-2018, both long and short-term estimates are obtained. This ARDL technique is used as it allows the cointegration of variables to check whether they are strictly integrated or in a mixed order. It is also suitable for tiny samples such as those used in this research (26 in this case). For the testing, the forest and/or F data are used. In a conditionally unconstrained ECM structure, the experiment examines the statistical significance of deficiencies in the model. Both the Wald / F statistics and the ECM offer an asymptotic distribution that is not standard and consistent (Pesaran, Shin, & Smith, 2001). The experiment examines the null hypothesis that the alternative theory of cointegration between factors is not being coincorporated. The use of the ECM is important because it is responsible for the degree to which previous innovations in the current period are corrected. Spurious regressions are avoided by applying the first variables that eliminate patterns in the variables. The ECM offers the most sophisticated model, too. The ECM's error correction expression is also stationary, which means that the error is not changed for a long time. The model of ARDL is as follows:

$$\Delta Y_t = \tau_0 + \sum_{i=1}^{k} \tau_1 \Delta Y_{t-i} + \sum_{i=0}^{k} \tau_n \Delta X_{t-i} + \lambda_1 Y_{t-1} + \lambda_n X_{t-i} + \varpi_t \qquad (3.5)$$

$Y_t$ and $X_t$ are respectively a dependent and an autonomous vector of variables. The first parameter of distinction is provided as $\Delta$ and $n$ the amount of regressors. The word mistake is specified as $\varpi$. The lag $k$ is evaluated in terms of duration. Dependent and vector lagged values of separate factors are specified as $Y_{t-i}$ and $X_{t-i}$ respectively. The short-term parameters of the autonomous variables vector are specified as $\tau_i$.

## 4. RESULTS AND DISCUSSIONS

Table 1.1 presents the unit root test results for the three variables.

**Table 1.1: Unit Root Test Results**

| Variable | ADF | | | | PP | | | |
|---|---|---|---|---|---|---|---|---|
| | Level | | First Difference | | Level | | First Difference | |
| | Trend | No Trend | Trend | No Trend | Trend | No Trend | Trend | No Trend |
| *LNINFL* | -0.903* | -0.144 | -0.439 | -0.125* | -0.903* | -2.681 | -0.856* | -4.264** |
| *LNINT* | -1.459 | -3.405** | -6.187** | _ | -2.444 | -3.374** | -5.228** | -4.301** |
| *LNOP* | -1.397 | -4.113** | --4.003** | _ | -1.432 | -3.933** | -3.830* | _ |

**Source:** Author
**Note:** (**)* implies statistical significance at 10 and 5 per cent levels respectively.

## Cointegration Test Results

Cointegration test results are given in Table 1.2. The ARDL (2, 1) and ARDL (1, 0,) are selected using the Schwarz Bayesian Criterion (SBC) for models 1 and 2.

**Table 1.2 ARDL bounds test for cointegration relationship**

| Test Statistic (F-statistic) | Model 1 | Model 2 |
| --- | --- | --- |
| | 45.5515** | 16.8296** |

**Source:** Author

**Note:** ***(*) indicates the null hypothesis of no cointegration is rejected at 5 per cent levels of statistical significance. The ARDL model gives the 95 per cent lower and upper bounds as 15.41-29.68.

The findings indicate that all models are cointegrated. The F-statistics for all designs are predicted to be higher than the critical upper bound value at a rate of 5%. The calculated F statistical is higher than the critical upper limit value at 5% of importance for Models 1 and 2. The variables are therefore actually co-integrated for all the models. The existence of a long-term partnership requires valid long-term and short-term outcomes. The research will then be carried out with the ARDL strategy to assess long to short term outcomes.

**3.5     Estimation Results**

**3.5.1 Estimated long-and short-run results for Model 1**

*3.5.1.1Long run results*

**Table 1.3: Estimated long-run coefficients for Model 1 using the ARDL approach**

| Variable | Coefficient | Std. Error | t-Statistic |
|----------|-------------|------------|-------------|
| *LNOP* | 3.340219* | 0.817971 | 4.083542 |
| *C* | 18.238330* | 3.526160 | 5.172292 |

**Source:** Author
**Note:** * implies statistical significance at 5 per cent levels of significance.

Table 1.3 presents the projected long-term outcomes of model 1.

At 5%, the crude oil price ratio is positive and statistically important. Inflation rises by roughly 3.34%, as the share of crude oil prices rises. This anticipated to lead to a rise in petroleum prices, and the company has to sell the products at greater market prices. Market price levels lead to a greater rate of inflation.  As oil is one of the most important factors of the manufacturing, an upward adjustment of the petroleum price has a direct impact on manufacturing costs and overall prices, leading to an increase in inflation.

This leads to greater inflation, as does crude oil prices. The economic surplus of petroleum income benefits Ghana. However, considering the exchange-trade volatility levels of Ghana, it contributes to longer-term, equal growth in Ghana.

**3.5.1.2 Short-run results**

Table 1.4 provides the short-term outcomes of model 1.

For model 1a the word error correction (ecm[-1]) of 1% is negatively and statistically important. It indicates that about 73% of shocks in the short run will be fixed in the long run. This implies that after a short-term shock, the long-term balance is adjusted very quickly. The adverse and statistically significant factor also confirms the derived and discussed co-integration relationship of Model 1 before.

**Table 1.4: Estimated short-run coefficients for Model 1 using the ARDL approach**

| Variable | Coefficient | Std. Error | t-Statistic |
|---|---|---|---|
| $\Delta LNINFL1$ | -0.117459 | 0.204460 | -0.574482 |
| $\Delta LNINFL2$ | -0.196539* | 0.100370 | -1.958153 |
| $\Delta LNOP$ | 0.533180* | 0.228610 | 2.332262 |
| $\Delta LNOP1$ | 0.453164* | 0.140083 | 3.234970 |
| $\Delta LNOP2$ | -0.346154 | 0.197005 | -1.757081 |
| $ECM(-1)$ | -0.730108** | 0.168967 | -4.321001 |
| $R-squared$ | 0.991714 | | |
| $Adj.R-squared$ | 0.978456 | | |
| $F-statistic$ | 74.80061** | | |
| $DW$ | 2.590319 | | |

**Source:** Author

**Note:** **(*) implies statistical significance at 1 and 5per cent levels of significance respectively.

The findings indicate that, in the short span (-0.117459), the coefficient of the initial inflation lag value is negative but statistically insignificant. However, the second-term inflation lag is positive and substantial statistically at a meaning point of 5 percent.

In short, the present value of the oil price and the laggings of one period is also positive and statistically important at a statistically significant rate of 5 percent. In particular, an rise in the value of crude oil by present and lagging for one period raises inflation by 0. Inflation. On a short-term basis, 53% and 0,45% respectively. The short-term results confirmed as above the long-term outcomes. But the lagged value of the crude oil price in two terms) (contradicts the long-term outcome, which is negatively and statistically insignificant.

The present results are in line with Ferraro, Rogoff and Rossi's (2012) work, who discovered that the cost of petroleum led to increased inflation. The results of Ahmed, Bhutto and Kalhoro (2019) show a positive correlation between crude oil prices and inflation, which has been confirmed by the present findings. An increase in manufacturing costs as a result of enhanced fuel costs is the direct impact of the crude oil price shock. The price of manufacturing would also increase, causing a decline in supply, as a consequence of inflation in the economy.

### 3.5.2 Estimation of Long-run and Short-run results for Model 2

Tables 1.5 and 1.6 show the projected long-term and short-run outcomes for models 2 and assess the impact of the interest rate on the crude oil prices, respectively.

### *3.5.2.1 Long-run results*

**Table 1.5: Estimated long-run coefficients for Model 2 using the ARDL approach**

| Variable | Coefficient | Std. Error | t-Statistic |
|---|---|---|---|
| *LNOP* | -3.968905* | 0.949932 | -4.178094 |
| *C* | 46.662061* | 8.351906 | 5.586995 |

**Source:** Author

**Note:** * implies statistical significance at 5 per cent levels of significance respectively.

The findings in the long run show that rising crude oil prices are leading to falling interest rates. This is because of the adverse and statistically important 5% level of importance of the coefficient of variable crude oil). This represents a 3,97% decrease in the interest rate for long-run crude oil prices. When interest rates fall, customers and businesses can loan and spend cash in a more free manner that boosts oil demand, from a theoretical view. The larger the use of oil, which has restrictions on manufacturing quantities imposed by OPEC, the higher the price for customers.

The Hotelling model believed that petroleum owners would decide whether or not to extract petroleum and sell it and leave it in the field, on the grounds of the true interest rate. If the cost of petroleum is growing so rapidly that it ensures greater returns than the selling cash, owners would prefer to keep it in the field–they will postpone their output to reach greater rates in the future (Hosek, Komarek & Motl, 2010). This decreases present supply and raises existing prices as future demand rises and future prices decrease.

### 3.5.2.2 Short-run results

Table 1.6 provides short-run outcomes for model 2.

The error correction ratio (ecm[-1]), at 1 percent of significance level, is negative and statistically important. After a shorter-term crude oil price shock, it demonstrates the rate of convergence towards long-term balance. From the outcomes 75.56% of shocks will be adapted in the long run if there is a short-term shock. Consequently, after a short-term shock, there is elevated rate of adjustment to long-term balance. The adverse and statistically important nature of ecm(-1) also supports the previously discussed co-integration relationship.

**Table 1.6: Estimated short-run coefficients for Model 2 using the ARDL approach**

| Variable | Coefficient | Std. Error | t-Statistic |
|----------|-------------|------------|-------------|
| $\Delta LNINT1$ | -0.106366 | 0.218658 | -0.486450 |
| $\Delta LNINT2$ | -0.300424* | 0.138968 | -2.161826 |
| $\Delta LNOP$ | 0.622341 | 0.915054 | 0.680114 |
| $\Delta LNOP1$ | -6.120987** | 1.336777 | -4.578914 |
| $\Delta LNOP2$ | 6.240736** | 1.513795 | 4.122577 |
| $ECM(-1)$ | -0.755771** | 0.268878 | -2.810831 |

**Source:** Author

**Note:** ** (*) implies statistical significance at 1 and per cent levels of significance respectively.

The short-term findings indicate that lagging interest rate values for the first and second period decrease existing interest rates. The two variables (and respectively) have an adverse coefficient. In particular, a first and second period lagging interest-rate increases in percentages lead to a reduction in present short-run interest rates of approximately 0,11 percent and 0.30 percent. However, only the short-term value of the interest rate for the second period is statistically important at 5 percent.

Existing rates) (and two periods of delayed value) (boost crude oil prices present rates of short-term interest, although only the latter are statistically important at 1%. A per cent upturn in present crude oil prices and two crude oil prices lagged about 0.62 and 6.24% respectively, short-term interest rises. The amount of importance is statistically important at 1 percent. The percentage increases of crude oil prices caused by the one-period decline in the interest rate in the short-term by the lagged value of crude oil price variable are roughly 6,12 percent.

If interest rates drop, all things being equal, there will be an impulse to slow down the present petroleum extraction and increase prices. The relationship between interest rates and oil prices should therefore be negative. If we combine the fast development in demand with the low actual interest rates seen in the last few years, a logical outcome of the Hotelling model is a quick growth in petroleum prices (Hosek, Komarek & Motl, 2010). The result of the research is consistent with the outcomes of Graces God and Maxwell (2013), concerning the connection between the cost of crude petroleum and the interest rate.

## 3.6    Model Adequacy and Reliability

Table 1.7 provides the results of a series of diagnostic and stability tests conducted on the ARDL model.

**Table 1.7 Model diagnostics and stability tests**

| Test statistic | Model 1 | Model 2b |
|----------------|---------|----------|
| Serial Correlation | 5.733585 | 0.731892 |
| | (0.0335) | (0.4974) |
| Functional Form | 0.123354 | 0.252726 |
| | (0.9049) | (0.8037) |
| Normality | 1.106782 | 0.198708 |
| | (0.574997) | (0.905422) |
| Heteroscedasticity | 0.531083 | 0.676298 |
| | (0.8751) | (0.7319) |
| CUSUM | Stable | Stable |
| CUSUMSQ | Stable | Stable |

**Source:** Author
**Note:** The values in parentheses are probability values.

The research examines serial and functional correlation by means of the Breusch-Godfrey and Ramsey Reset test. The Breusch-Pagan-Godfrey test also examines the heteroscedasticity when the Jarque-Berra test determines normality. Further CUSUM and CUSUMSQ tests are employed to determine structural stability. All models are diagnosed with the model and stability tests performed based on the outcomes provided in Table 4.7. All models have no serial correlation, functional shape, normality and heteroscedasticity. The findings achieved and discussed therefore reliable for all models.

## 5. Conclusions

The study used monthly scheduling data from 2000 to 2018 for the macroeconomic effect of crude oil price in Ghana.

The findings of Model 1 indicate that crude oil price and long-term inflation have a favorable connection. Inflation is greater because of the price level on the market. The short duration shows an adverse but statistically insignificant coefficient of inflation of its first lag in the short term. However, inflation is positive and substantial in the second laggard era. The value of oil price in the short term is positive and significant, both now and for a period of time. With inflation in the short-term rising in the current percentage and crude oil value lagging at only one period. Two periods of negative and insignificant crude oil price values are at odds with the long-term effect.

For Model 2, a rise in the cost of crude oil led to a drop in the rate of interest. Thus, customers and businesses can borrow and spend cash freer, which boosts demand for oil, when interest rates fall. The short-term findings show that interest-rate values lagging in the first and second period decrease present rate rates. The two variables have an adverse coefficient. A first and second period rise in lag period for crude oil prices leads to a decline in existing short-term interest rates. The present level of crude oil prices is up by a proportion and the two lagging crude oil prices are causing short-term interest rises. A single-period rise in the value of crude oil could lead to a decrease in the short-run rate of interest.

**REFERENCES**

Abouali, O., Nikbakht, A., Ahmadi, G., & Saadabadi, S. (2009). Three-dimensional simulation of Brownian motion of nano-particles in aerodynamic lenses. *Aerosol Science and Technology, 43*(3), 205-215.

Acheampong, I. (2007). Testing Mckinnon-Shaw thesis in the context of Ghana's financial sector liberalisation episode. *International Journal of Management Research and Technology, 1*(2), 156-183.

Adeniyi, O., Oyinlola, A., & Omisakin, O. (2011). Oil price shocks and economic growth in Nigeria: are thresholds important? *OPEC Energy Review, 35*(4), 308-333.

Ahmed, K., Bhutto, N. A., & Kalhoro, M. R. (2019). Decomposing the links between oil price shocks and macroeconomic indicators: Evidence from SAARC region. *Resources Policy, 61*, 423-432.

Bala, U., & Chin, L. (2018). Asymmetric impacts of oil price on inflation: An empirical study of African OPEC member countries. *Energies, 11*(11), 3017.

Basnet, H. C., & Upadhyaya, K. P. (2015). Impact of oil price shocks on output, inflation and the real exchange rate: evidence from selected ASEAN countries. *Applied Economics, 47*(29), 3078-3091.

Bawumia, M., & Halland, H. (2017). Oil discovery and macroeconomic management. *Extractive Industries*, 220.

Creswell, J. W., & Poth, C. N. (2017). *Qualitative inquiry and research design: Choosing among five approaches*: Sage publications.

Ekmekcioglu, E. (2012). The macroeconomic effects of world crude oil price changes. *International Journal of Business and Social Science, 3*(6).

Ewing, B. T., & Malik, F. (2017). Modelling asymmetric volatility in oil prices under structural breaks. *Energy Economics, 63*, 227-233.

Ferraro, D., Rogoff, K. S., & Rossi, B. (2012). *Can oil prices forecast exchange rates?* Retrieved from

Foresti, P. (2006). Testing for Granger causality between stock prices and economic growth.

Gatsi, J. G., Gadzo, S. G., Anipa, C. A. A., & Kosipa, S. (2015). Empirical analysis of macroeconomic factors and crude oil prices in Ghana. *International Journal of Economics, Commerce and Management, 3*(9), 217-235.

Gay, R. D. (2016). Effect of macroeconomic variables on stock market returns for four emerging economies: Brazil, Russia, India, and China. *International Business & Economics Research Journal (IBER), 15*(3), 119-126.

Gorard, S., & Gorard, J. (2016). What to do instead of significance testing? Calculating the 'number of counterfactual cases needed to disturb a finding'. *International Journal of Social Research Methodology, 19*(4), 481-490.

Hosek, J., Komarek, L., & Motl, M. (2010). *Monetary Policy and Oil Prices*. Retrieved from

Hussain, M., Zebende, G. F., Bashir, U., & Donghong, D. (2017). Oil price and exchange rate co-movements in Asian countries: Detrended cross-correlation approach. *Physica A: Statistical Mechanics and its Applications, 465*, 338-346.

Hye, Q. M. A., & Riaz, S. (2008). Causality between Energy Consumption and Economic Growth.

Jacobs, J., Kuper, G. H., & Van Soest, D. P. (2009). On the effect of high energy prices on investment. *Applied Economics, 41*(27), 3483-3490.

Kang, W., Ratti, R. A., & Yoon, K. H. (2015). The impact of oil price shocks on the stock market return and volatility relationship. *Journal of International Financial Markets, Institutions and Money, 34*, 41-54.

Kibunyi, A., Nzai, C. C., & Wanjala, K. (2018). Effect of Crude Oil Prices on GDP Growth and Selected Macroeconomic Variables in Kenya. *Journal of Economics and Business, 1*(3), 282-298.

Kilian, L., & Hicks, B. (2013). Did unexpectedly strong economic growth cause the oil price shock of 2003–2008? *Journal of Forecasting, 32*(5), 385-394.

Kilian, L., & Vigfusson, R. J. (2011). Are the responses of the US economy asymmetric in energy price increases and decreases? *Quantitative Economics, 2*(3), 419-453.

Kotini, A. G., Chang, C.-J., Chow, A., Yuan, H., Ho, T.-C., Wang, T., . . . Olszewska, M. (2017). Stage-specific human induced pluripotent stem cells map the progression of myeloid transformation to transplantable leukemia. *Cell Stem Cell, 20*(3), 315-328. e317.

MoF. (2017). *2018 Government of Ghana Budget Statement and Economic Planning* (657). Ghana: Ministry of Finance

Omisakin, D., & Olusegun, A. (2008). Oil price shocks and the Nigerian economy: a forecast error variance decomposition analysis. *Journal of Economic Theory, 2*(4), 124-130.

Pesaran, M. H., Shin, Y., & Smith, R. J. (2001). Bounds testing approaches to the analysis of level relationships. *Journal of applied econometrics, 16*(3), 289-326.

Robson, M. T. (1993). Federal funding and the level of private expenditure on basic research. *Southern Economic Journal*, 63-71.

Taiwo, M., Abayomi, T., & Damilare, O. (2012). Crude oil price, stock price and some selected macroeconomic indicators: Implications on the growth of Nigeria economy. *Research Journal of Finance and Accounting, 3*(2), 42-48.

ThankGod, A. O., & Maxwell, I. A. (2013). Macroeconomic impact of oil price levels and volatility in Nigeria. *International Journal of Academic research in Economics and Management sciences, 2*(4), 15.

Trochim, W. (2000). Research methods knowledge base: Survey research. *Am J Sports Med, 30*(6), 212.

Wilson, A., David, U., Inyiama, O., & Beatrice, E. (2014). Oil price volatility and economic development: Stylized evidence in Nigeria. *Journal of Economics and International Finance, 6*(6), 125-133.

Zhihua, D., Meihua, Z., & Bo, N. (2011). Research on the influencing effect of coal price fluctuation on CPI of China. *Energy Procedia, 5*, 1508-1513.

Zhu, H.-M., Li, R., & Li, S. (2014). Modelling dynamic dependence between crude oil prices and Asia-Pacific stock market returns. *International Review of Economics & Finance, 29*, 208-223.